# Digitally Supported Transitional Care Pathways for Older Adults: A Multi-Site Co-Design Study

**Camilla Gabrielsen[1], Roberta Bevilacqua[2], Lorena Rossi[2], Eleonora Bonifazi[2], Margherita Osimani[2], Ovidiu Anchidin[3], Ancuta Nemes[3], Ionut Anghel[*4], Tudor Cioara[4], Terje Grimstad[5], Riitta Hellman[5], Arnor Solberg[6] and Lars Thomas Boye[6]**

[1] Farsund Kommune, Postboks 100, 4552 Farsund, Norway
[2] Scientific Direction, IRCCS INRCA, Via S. Margherita 5, 60124 Ancona, Italy
[3] Niculae Stancioiu Heart Institute, Moților 19-21, 400001 Cluj-Napoca, Romania
[4] Computer Science Department, Technical University of Cluj-Napoca, G. Barițiu 26-28, 400027 Cluj-Napoca, Romania
[5] Karde AS, Irisveien 14, 0870 Oslo, Norway
[6] Tellu AS, Lysaker Torg 15, 1366 Lysaker, Norway
*Corresponding Author: ionut.anghel@cs.utcluj.ro;

**Abstract.**

**Background:** During transitions from hospital to home, older adults are frequently affected by fragmented communication, limited coordination between levels of care, and insufficient post-discharge follow-up. Digital technologies, including remote patient monitoring (RPM), digital communication, and artificial intelligence (AI), offer opportunities to improve continuity of care. However, their integration needs to reflect the organizational and clinical contexts of different healthcare systems. This study presents the development and early implementation of digitally supported transitional care pathways within the TransCare project.

**Methods:** Care pathways were analyzed and redesigned across three pilot sites in Italy, Romania, and Norway. Existing clinical and organizational processes were mapped to identify challenges and opportunities for improvement. Technology-augmented pathway matrices were developed to integrate RPM, digital communication tools, and AI-supported analytics across the stages of assessment, hospitalization, discharge, and post-discharge follow-up. The pathways were iteratively refined through a co-design process involving healthcare professionals, patients, caregivers, and technical partners. Early implementation experiences were subsequently analyzed across the three sites.

**Results:** Common challenges included fragmented communication and poor care coordination, particularly during discharge and post-discharge follow-up. However, the underlying cause and required solutions differed across the pilot sites, reflecting variations in patient profiles, healthcare organizations, available resources, and digital maturity. The redesigned pathways improved role clarity, strengthened discharge planning, and enabled more systematic follow-up. Early implementation experience also highlighted the importance of patient and caregiver preparation, professional training, organizational readiness, and adaptable digital solutions.

**Conclusion:** Digitally supported transitional care requires the integration of technology with organizational and clinical process redesign. Future work will focus on testing the redesigned pathways in real-world pilot settings and evaluating their feasibility and impact on transitional care outcomes and patient experience.

**Keywords:** Care Pathways, Transitional Care, Rehospitalization rate, Artificial Intelligence, Remote patient monitoring.

## 1. Introduction

Population ageing is placing sustained pressure on European healthcare systems, particularly at the point where patients transition from hospital to home [1]. This transition, commonly referred to as transitional care, is a well-documented weak point in the care continuum, especially for frail older adults with chronic, multimorbid conditions [2]. At the same time older adults' hospital readmissions are a large-scale, cross-cutting problem rather than one associated with a specific disease [3]. Longitudinal evidence indicates that roughly one in five older adults receiving home healthcare services is readmitted within 30 days of discharge. The risk of readmission varies with the reason for the original hospitalization, with infections and other acute conditions among older populations [4]. Moreover, avoidable readmissions vary widely, depending on the setting and methodology used, underscoring that the causes are heterogeneous rather than tied to any single pathology. These findings suggest a broader structural problem rather than isolated clinical failures, including: discontinuity of information, fragmented responsibility between hospital and community-based providers, and insufficient post-discharge follow-up mechanisms.

The role of digitally supported transitional care and telemedicine in mitigating readmissions has received increasing attention in recent years [5]. Modern remote patient monitoring (RPM) systems can integrate AI-based analytics to support proactive, data-driven transitional care. By detecting early signs of clinical deterioration from remotely monitored patient data, these systems can shift follow-up from reactive to anticipatory intervention [6]. AI tools in transitional care also aim to enhance transitions from hospital to home by improving discharge planning, follow-up care, and interoperability. Care continuity and complexity management are among the most frequently targeted areas, while patient and caregiver engagement remain comparatively underserved [7]. Evidence from RPM and telemonitoring trials further suggests that digitally enabled follow-up can reduce avoidable hospitalizations when appropriately integrated into clinical workflows [8]. However, technology adoption alone is insufficient without accompanying redesign of roles, heterogeneous care pathways, communication channels, and reimbursement structures. At the same time, research has largely focused on evaluating individual technologies in isolation or proposing generic frameworks that have not been validated through comparative, multi-country implementation. This gap in translating pathway-level redesign methodologies into pilot-validated, context-adapted implementations motivates the present work.

This redesign process, however, presents several important challenges. First, technology integration must be adapted to the digital maturity, infrastructure, and organizational context of each care setting. A single standardized pathway cannot simply be transferred across healthcare systems with different levels of resourcing, interoperability, and regulatory constraints [9]. Second, successful redesign depends on clearly defined roles and responsibilities among the many actors involved, including patients, family caregivers, general practitioners (GPs), hospital teams, and community or municipal services. Ambiguity at key handover points, particularly during discharge, is where continuity of care most often breaks down [10]. Third, adequate training and preparation of both patients and healthcare professionals before discharge are essential. The benefits of RPM and digital communication tools depend heavily on user engagement and correct use, particularly among frail older adults and individuals with cognitive impairment [11]. Fourth, redesigned pathways must be validated through iterative real-world piloting rather than assumed to be effective at the design stage. Early implementation frequently reveals discrepancies between intended workflows and actual clinical practice, highlighting the need for continuous refinement [12]. Finally, sustainability and scalability remain important concerns. Financial models, data protection requirements, and interoperability between digital platforms and healthcare organizations must be addressed before technology-enabled pathways can move beyond pilot-stage validation and into routine practice [13].

Simply layering digital tools onto existing transitional care processes is insufficient to improve continuity of care [14]. Technology alone does not resolve coordination challenges; rather, it acts as an enabler within a well-structured care pathway. Redesigning care pathways to explicitly integrate RPM, AI-based decision support, and digital communication platforms is therefore a necessary step. Such a redesign enables care teams to move from reactive, episodic contact to proactive, data-driven follow-up. It embeds patient screening and technology onboarding before discharge, continuous monitoring and alert-based interventions after discharge, and scheduled follow-up at defined intervals directly into the care pathway rather than treating these activities as add-ons.

TransCare (New care pathways for supporting transitional care from hospitals to home using AI and personalized digital assistance) is an EU-funded project under the Transforming Health and Care Systems (THCS) programme [15]. The project brings together a consortium of clinical, technological, and municipal partners across three countries: Italy, Romania, and Norway. Its overarching goal is to design, implement, and clinically validate a technology-assisted transitional care model that integrates RPM, Artificial Intelligence (AI)-based decision support, and digital communication platforms into redesigned, patient-centered care pathways. The model aims to improve continuity of care, reduce hospital readmissions, and ease the burden on healthcare providers. The pilot sites are comprised of northern, eastern, and southern EU and cover the entire care continuum. Istituto di Ricovero e Cura a Carattere Scientifico IRCCS-INRCA in Italy is a geriatric clinical and research hospital, and it is the leader of Italian Ageing Network. They will play an important role in transitional care by providing support and services to individuals who are transitioning from one healthcare setting to another, bridging the gap between acute care and long-term care or home-based care. The Heart Institute Niculae Stăncioiu (HINS) is one of the most important tertiary cardiology centers in Romania. This pilot will play a key role in conducting clinical trials by identifying suitable candidates from their inpatient population who meet the defined eligibility criteria and evaluating and refining transitional care practices for cardiovascular diseases. Farsund municipality in Norway (FAR) is a community-based care organization supporting rehabilitation and independent living. The objective is to support the transition of these patients to receive care at home by implementing specific municipality health services using the technology developed in the project.

The main contributions of this paper are:

- to identify key challenges in transitional care and analyse existing care pathways across the three TransCare pilot countries;
- to re-design technology-augmented care pathway matrices that demonstrate how RPM, AI, and digital communication can be integrated into each phase of the patient journey, from assessment and admission through hospital stay, discharge, and post-discharge follow-up, while preserving the distinct clinical and organizational context of different countries.;
- to discuss early findings and derive cross-site lessons learned, providing transferable insights for the design of technology-enabled transitional care pathways in European healthcare systems.

The remainder of this paper is structured as follows: Section 2 presents the methods used for transforming traditional care pathways into technology enabled Pathways. Section 3 presents the redesigned care pathways. Section 4 discusses the findings and lessons learned. Finally, Section 5 concludes the paper.

## 2. Methods

### 2.1. National challenges in transitional care

It is well recognized that transitions from hospital to municipal care is challenging due to communication gaps between healthcare providers, inadequate discharge planning, resource limitations in community settings, and the need for effective interprofessional collaboration. Additionally, patients and their families often feel unprepared for the transition, leading to anxiety and stress. Addressing these challenges requires health and social care providers to work together, with shared focus on effective discharge planning and smoother care transitions. Within the TransCare project, the three pilot countries (Italy, Romania and Norway) conducted interviews with patients, GPs, hospital staff, and municipality healthcare professionals to identify local challenges in transitional care. Workshops were also conducted to obtain an interdisciplinary perspective on transitional care. The main challenges identified across the three pilot countries are presented below.

In Italy, the main challenges in transitional care identified within the TransCare project include regional disparities, lack of standardized discharge protocols, financial constraints, and limited integration between hospital and community services [16-18]. Regional differences in the provision of community care services remain substantial. While some regions have invested heavily in community-based services and are close to meeting national targets, others are still in the early stages of development. In addition, funding allocated through the National Recovery and Resilience Plan (PNRR) was established before the recent period of inflation and primarily covers capital expenditures, with limited provision for increased costs related to personnel, medical devices, and medications. Difficulties in recruiting healthcare professionals represent an additional challenge. These difficulties are linked to financial constraints and to resistance among some GPs, who are concerned that a shift toward community-based care may reduce their professional autonomy. As a result, integration between hospital and community services remains limited. This challenge is further compounded by the need to modernize digital infrastructure and accelerate the digital transition. Digitalization must also address the needs of an ageing population, many of whom may be unfamiliar with or reluctant to use digital services. Furthermore, the discharge of older adults with complex and often chronic medical conditions can make effective management at home particularly challenging.

In Romania, the healthcare system faces significant challenges in transitional care. Most patients are discharged to home following treatment in tertiary care centers. In some cases, patients do not have close relatives and are unable to care for themselves independently, requiring healthcare professionals to involve social services. According to current legislation, people with severe disabilities may receive social support services upon request or through an ex officio decision, including access to a caregiver or personal assistant. The role may be assumed by a family member, who receives financial compensation for providing care. However, there is a severe shortage of social workers funded through the public healthcare and social support system, limiting access to these services. According to Eurostat, the Romanian social assistance system is among the least effective in the European Union [19]. In this context, digital health solutions may offer support, but they require a level of digital competence that many patients do not possess. Communication and information sharing between healthcare providers remain limited. In many cases, medical information is not exchanged effectively between care providers, and the management and transfer of medical documentation continue to represent significant challenges within the Romanian healthcare system. Although the GP is theoretically the healthcare professional with the most complete overview of the patient's medical history, this role is not always fully utilized. Greater involvement of GPs in transitional care could strengthen coordination between specialist care, social services, and the patient's family.

Regional disparities present an additional challenge, as access to healthcare services remains limited in many rural areas. GPs, particularly in these regions, are often overburdened, manage long patient waiting lists, and have limited capacity to engage in additional aspects of patient care and follow-up [20-21].

In Norway, one of the main challenges in transitional care is poor coordination between municipal healthcare services and hospitals. This disconnect often leads to fragmented care and unclear responsibilities for patient follow-up after discharge. Older adults are particularly affected by these coordination challenges. The National Health and Hospital Plan has sought to address this issue through the establishment of health communities aimed at strengthening cooperation between hospitals and municipal services [22]. Another pressing challenge is the shortage of healthcare professionals, particularly within municipal health services. A lack of trained personnel makes it difficult to provide continuous and effective care transitions, including patient transfers and post-discharge follow-up. According to the National Health and Hospital Plan, workforce shortages reduce the system's ability to meet patient needs and increase pressure on existing staff, potentially affecting the quality of care. The absence of standardized procedures further contributes to inconsistent care during transitions. Municipalities and hospitals often rely on local routines and practices, creating communication gaps and delays in treatment. As a result, the lack of uniform care pathways remains a significant barrier to effective transitional care and highlights the need for improved integration across healthcare services [23]. Patient and family involvement represents another important challenge. Many patients and their families feel excluded from decision-making during care transitions, leaving them insufficiently prepared to manage care after discharge. This may contribute to avoidable readmissions and poorer health outcomes. Involving next of kin as active partners in the transition process can improve both communication and quality of care [24]. The integration of technology into transitional care also remains limited. Despite the potential of digital solutions to improve communication and coordination across care levels, many healthcare providers continue to rely on fragmented information systems, resulting in insufficient information sharing. Follow-up after hospital discharge is often inadequate, increasing the risk of complications, readmissions, and patient dissatisfaction, particularly among individuals with chronic illnesses or complex care needs. Consequently, insufficient follow-up remains a systemic challenge, largely driven by poor coordination between primary and specialist healthcare services.

Despite their contextual differences, all three sites converge on the same underlying problem: transitional care pathways that were never explicitly designed to support structured information exchange, clearly defined responsibilities, and technology-enabled follow-up across the hospital-to-home continuum. Therefore, ensuring continuity of care requires multidimensional assessment, effective communication between hospital and community services, standardized care processes, and systematic sharing of patient information.

### 2.2. Analytical framework

The methodology for redesigning transitional care pathways in TransCare is based on a structured, evidence-informed framework that combines qualitative insights with process mapping (see Figure 1). The analytical framework comprises three interrelated activities:

- *Mapping current practices* through care pathway matrices that visualize the patient journey from hospital to home.
- *Identification of systemic and local challenges* using stakeholder interviews and interdisciplinary workshops.
- *Iterative co-design* to develop improved pathways that integrate digital tools and reflect real-world constraints and opportunities.

This framework ensures that the redesigned pathways are both context-sensitive and aligned with the technological capabilities developed in TransCare. The redesign process is deeply participatory, involving a wide range of stakeholders across all pilot sites:

- *Patients and caregivers* were interviewed to understand their experiences, needs, and expectations during care transitions.
- *GPs, hospital clinicians, and municipal care providers* contributed insights into current workflows, coordination challenges, and opportunities for improvement.
- *Multidisciplinary teams* participated in workshops to collaboratively map existing pathways and identify opportunities for improvement.
- *Technology partners and researchers* supported the integration of digital solutions and ensured methodological rigor.

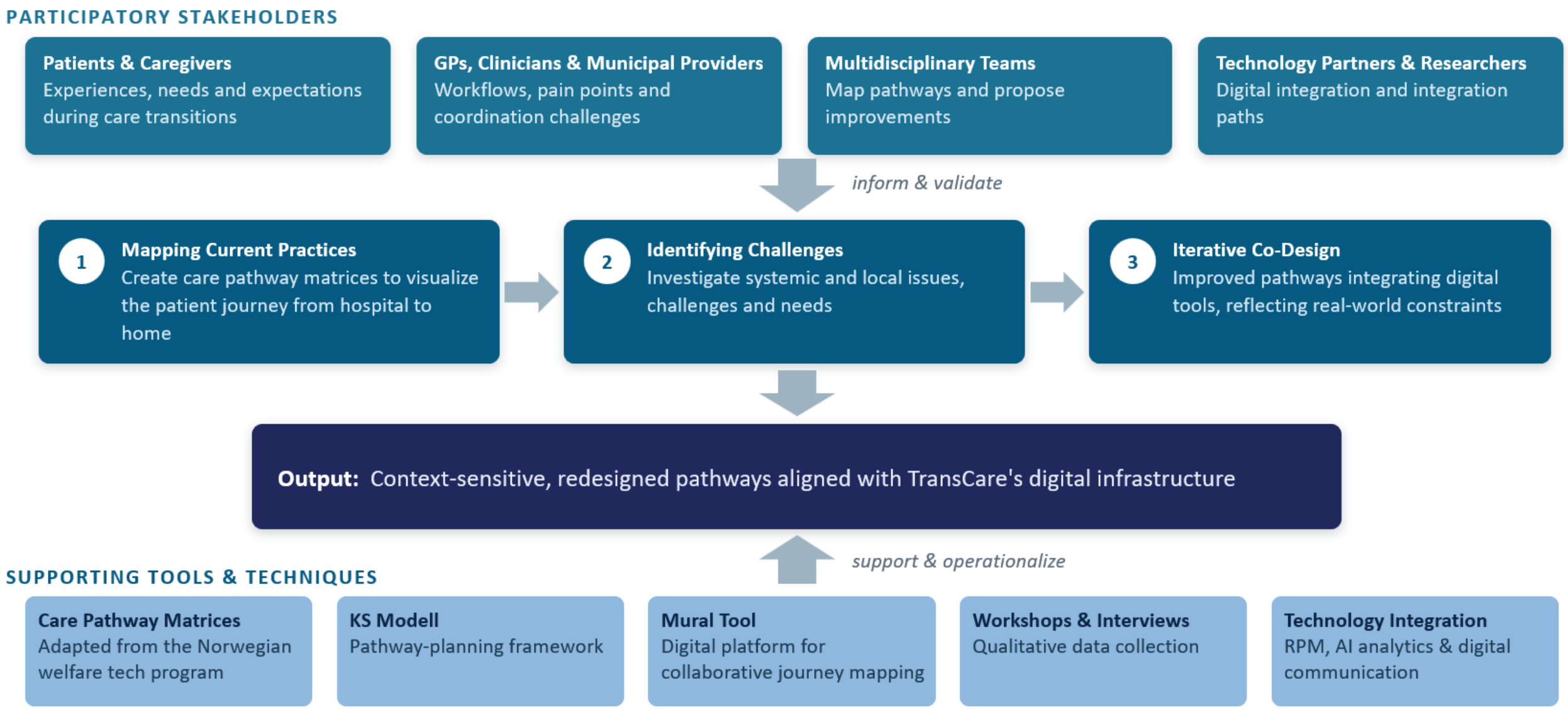

**Figure 1: Analytical framework for redefining transitional care pathways.**

This co-design approach promotes ownership, relevance, and feasibility of the redesigned pathways, ensuring that they are tailored to the specific context of each healthcare setting. Several tools and techniques were employed to support the mapping, analysis, and redesign of care pathways:

- *Care Pathway Matrices:* Originally developed by PA Consulting for the Norwegian welfare technology program [25] these matrices were adapted for each pilot site to visualize care process, stakeholder roles, and information flows. They served as diagnostic tools to identify inefficiencies and fragmentation and opportunities for improvement.
- *KS Modell*: A structured framework [26] used to guide patient pathway planning, emphasizing assessment, data sharing, patient engagement, and follow-up. The framework was adapted to reflect the specific organizational and healthcare context for each country.
- *Mural Tool*: A digital visualization platform [27] used to collaboratively map patient journeys and identify challenges within the care process. The plattform supported stakeholder engagement and facilitated the visualization of complex care transitions.
- *Workshops and Interviews*: Conducted across all sites to gather qualitative data and validate findings. These activities ensured that the redesign process was informed by the experiences and perspectives of those directly involved in delivering and receiving care.
- *Technology integration*: The methodology is closely aligned with the digital infrastructure developed in TransCare [15], ensuring that redesigned pathways are compatible with RPM, AI-based analytics, and digital communication tools.

This multi-method approach ensures that the redesigned pathways are not only clinically sound, but also operationally feasible and technologically integrated.

### 2.3. Design of the care pathway matrices

The initial care pathway matrices were developed using principles derived from the KS Modell and adapted for the TransCare context to systematically examine current transitional care practices, with particular emphasis on communication and discharge planning. The matrices map the patient journey from home to hospital and back to the community, highlighting critical transition points where coordination between hospitals and municipal services is essential. They also identify common challenges, including delayed or missing notifications to municipalities, unclear or inconsistent documentation, and insufficient collaboration during discharge planning. By visualizing these processes and challenges, the matrices serve as a tool for identifying gaps and improving continuity and quality of care across healthcare settings.

Figure 2 illustrates the current patient journey at INRCA across four main phases: (1) assessment and clarification, (2) hospital admission, (3) hospital stay, and (4) discharge. During the assessment phase, patients seek medical attention following an illness or deterioration in their condition. Family members are often involved in supporting communication with healthcare professionals and facilitating access to care. GPs assess the patient and may refer them to specialist services, while home healthcare services report emergencies when required. During hospital admission, patients undergo admission procedures and a comprehensive geriatric assessment. Family members contribute relevant clinical and medication information, while home healthcare services suspend ongoing services and update patient records as appropriate.  The hospital stay phase involves comprehensive geriatric assessment, monitoring of nutritional and cognitive status, and rehabilitation when necessary. Care is coordinated through collaboration among physicians, nurses, and rehabilitation professionals, with ongoing communication regarding the patient's progress and care plan.  At discharge, the hospital develops a therapeutic plan, prepares discharge documentation, and coordinates follow-up services. Home healthcare services may resume care following discharge, including assessments and ongoing monitoring. Throughout the pathway, family members play an important role in supporting communication, coordinating care, and facilitating the transition from hospital to home, particularly for patients with reduced functional independence.  Overall, the pathway demonstrates the complexity of transitional care and the need for effective communication, coordination, and information sharing among multiple stakeholders to ensure continuity of care.

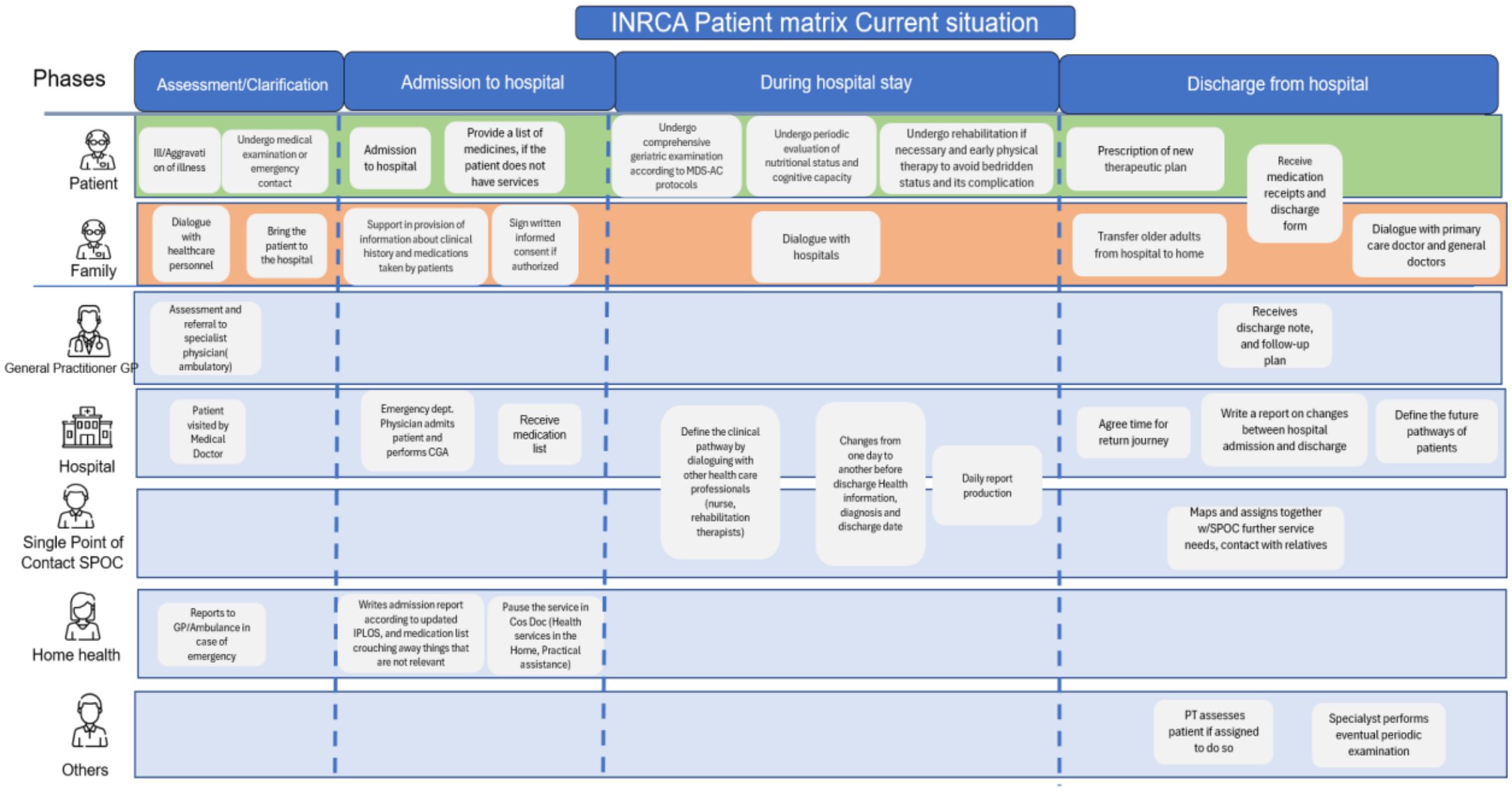

**Figure 2: Italy specific care pathway.**

Figure 3 presents the current patient journey at FAR across five key phases: (1) assessment and clarification, (2) hospital admission, (3) hospital stay, (4) discharge from hospital, and (5) post-discharge follow-up. During the assessment phase, patients seek healthcare services following the onset or worsening of symptoms, often with support from family members. GPs assess the patient and may refer them to hospital services, while coordination between hospital and municipal healthcare services is initiated when necessary.  During hospital admission and the subsequent hospital stay, patients receive diagnostic assessment, treatment, and monitoring. Care is coordinated by a multidisciplinary hospital team, and communication with patients and family members is maintained throughout the admission. Discharge planning is initiated early in the hospital stay and includes assessment of post-discharge care needs and coordination with municipal healthcare services. Discharge requires collaboration between hospital staff, the Single Point of Contact (SPOC), municipal healthcare providers, patients, and family members. Discharge documentation, medication information, and arrangements for ongoing care are prepared to support a safe transition home. Required assistive devices, home modifications, and support services are coordinated before discharge whenever possible. Following discharge, patients continue their recovery at home with support from family members, GPs, and municipal healthcare services. Ongoing follow-up and coordination aim to promote recovery, maintain continuity of care, and reduce the risk of hospital readmission. Overall, the pathway highlights the importance of early discharge planning, multidisciplinary collaboration, and effective coordination between hospital and municipal healthcare services to ensure safe and seamless care transitions.

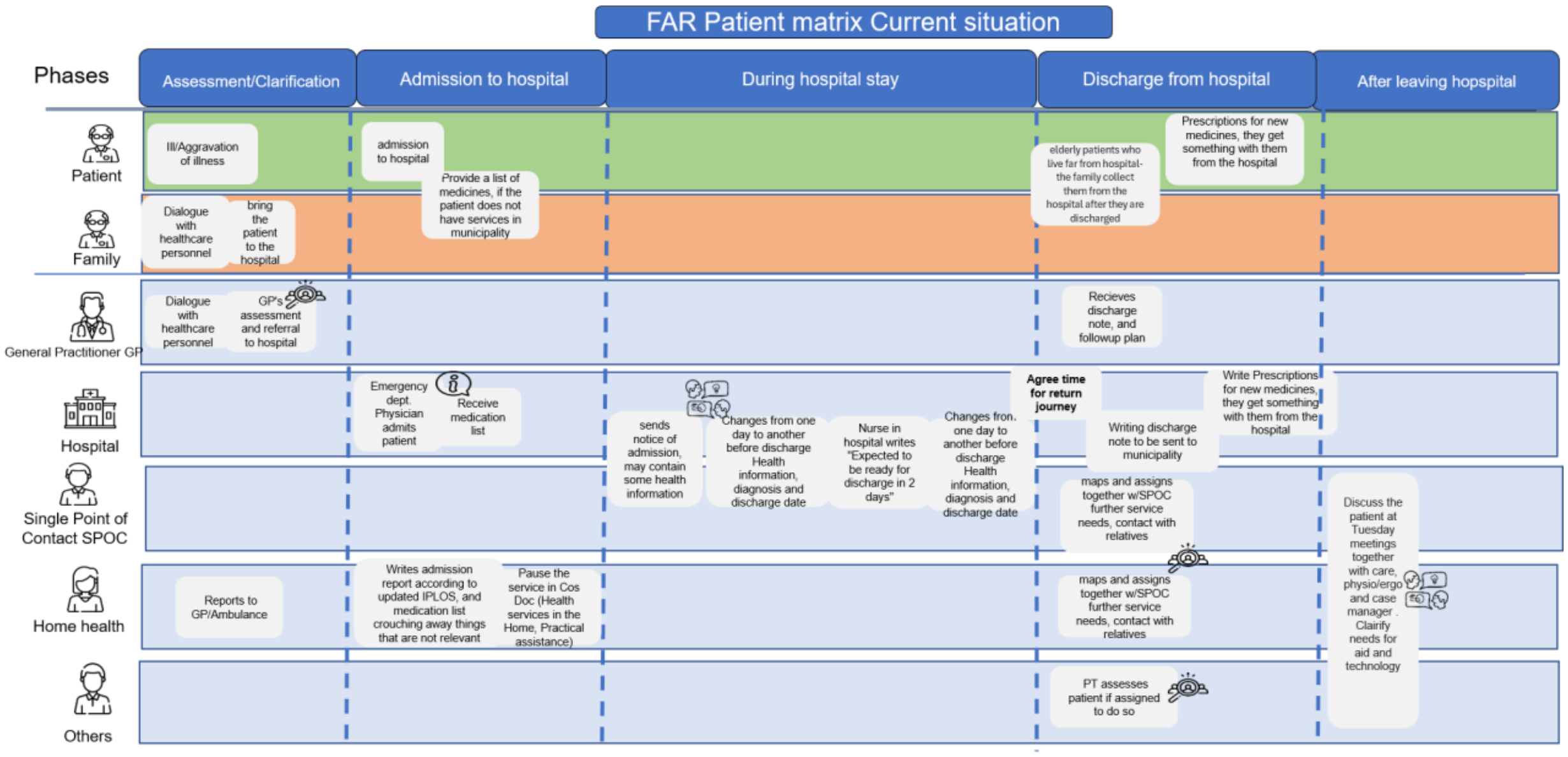

**Figure 3: Norway specific care pathway.**

Figure 4 presents the current patient journey at HINS across five key phases: (1) assessment and clarification, (2) hospital admission, (3) hospital stay, (4) discharge from hospital, and (5) post-discharge follow-up. During the assessment phase, patients seek medical attention following the onset of new symptoms or deterioration of an existing condition. In most cases, patients are initially assessed by their GP and subsequently referred to an outpatient cardiologist, who determines whether outpatient management or hospital admission is required. Patients with acute or severe symptoms may present directly to the emergency department or arrive via emergency medical services. Family members frequently support the patient during this stage, particularly with communication and transportation. Upon hospital admission, patients undergo a comprehensive clinical assessment, and family members may provide relevant medical information when necessary. During the hospital stay, additional investigations are performed to establish a final diagnosis and guide for treatment. Depending on the patient's condition, care may include specialized cardiological treatment, invasive procedures, or cardiac surgery. Throughout admission, patients are clinically monitored and informed about their condition, treatment plan, and anticipated discharge. At discharge, patients receive a discharge summary, treatment recommendations, and a follow-up plan. Coordination of post-discharge care involves both GPs and outpatient cardiologists, who are responsible for continued monitoring and treatment according to the discharge recommendations. Overall, the pathway highlights the importance of coordination between primary care, specialist cardiology services, and hospital-based care to support continuity of care and effective follow-up after discharge.

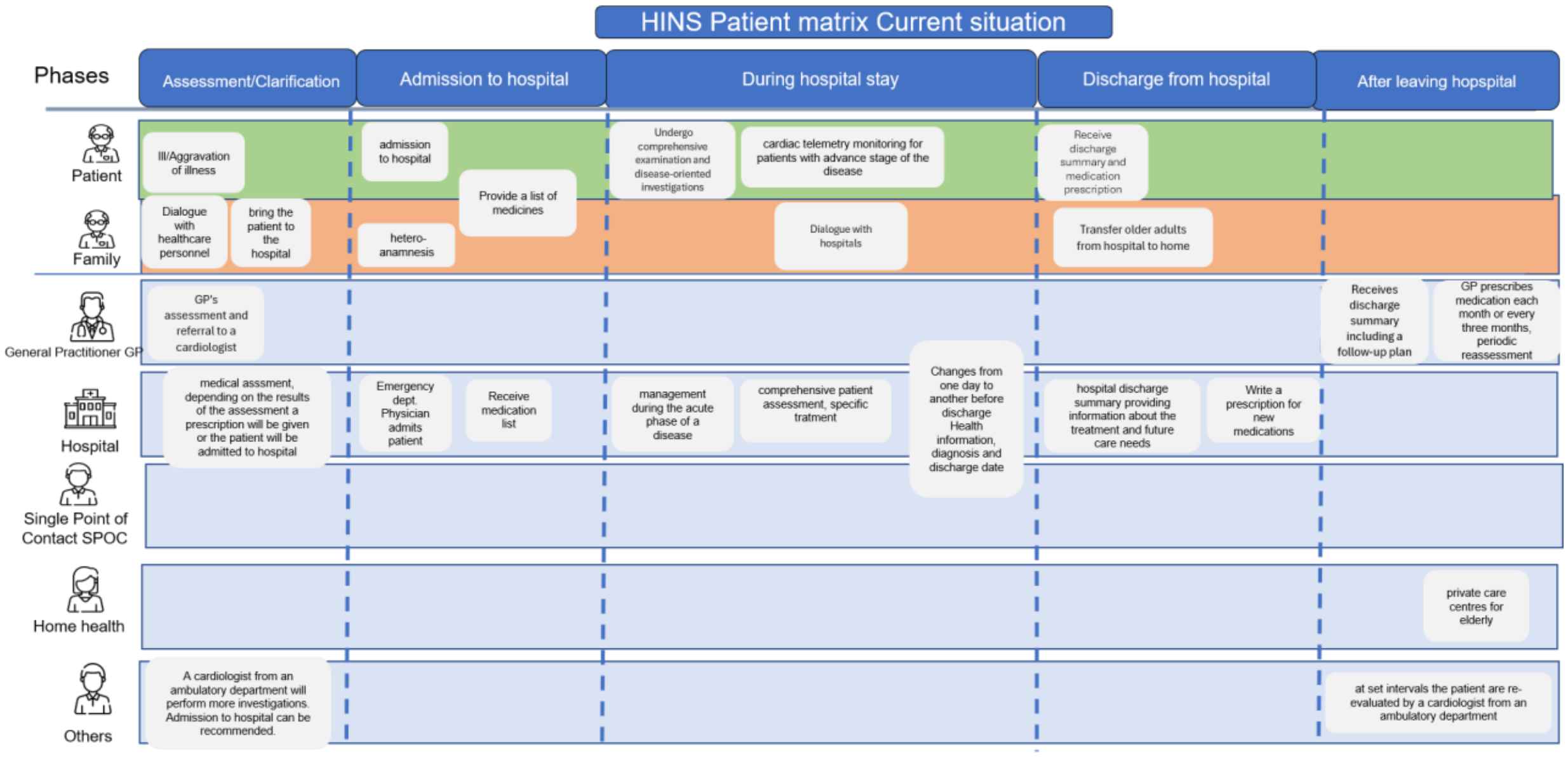

**Figure 4: HINS specific care pathway.**

### 2.4. Technology integration

The TransCare platform integrates a range of digital technologies designed to support older adults during the transition from hospital to home. These technologies aim to enhance patient self-management, strengthen communication across care settings, and support continuity of care while reducing the risk of hospital readmission.

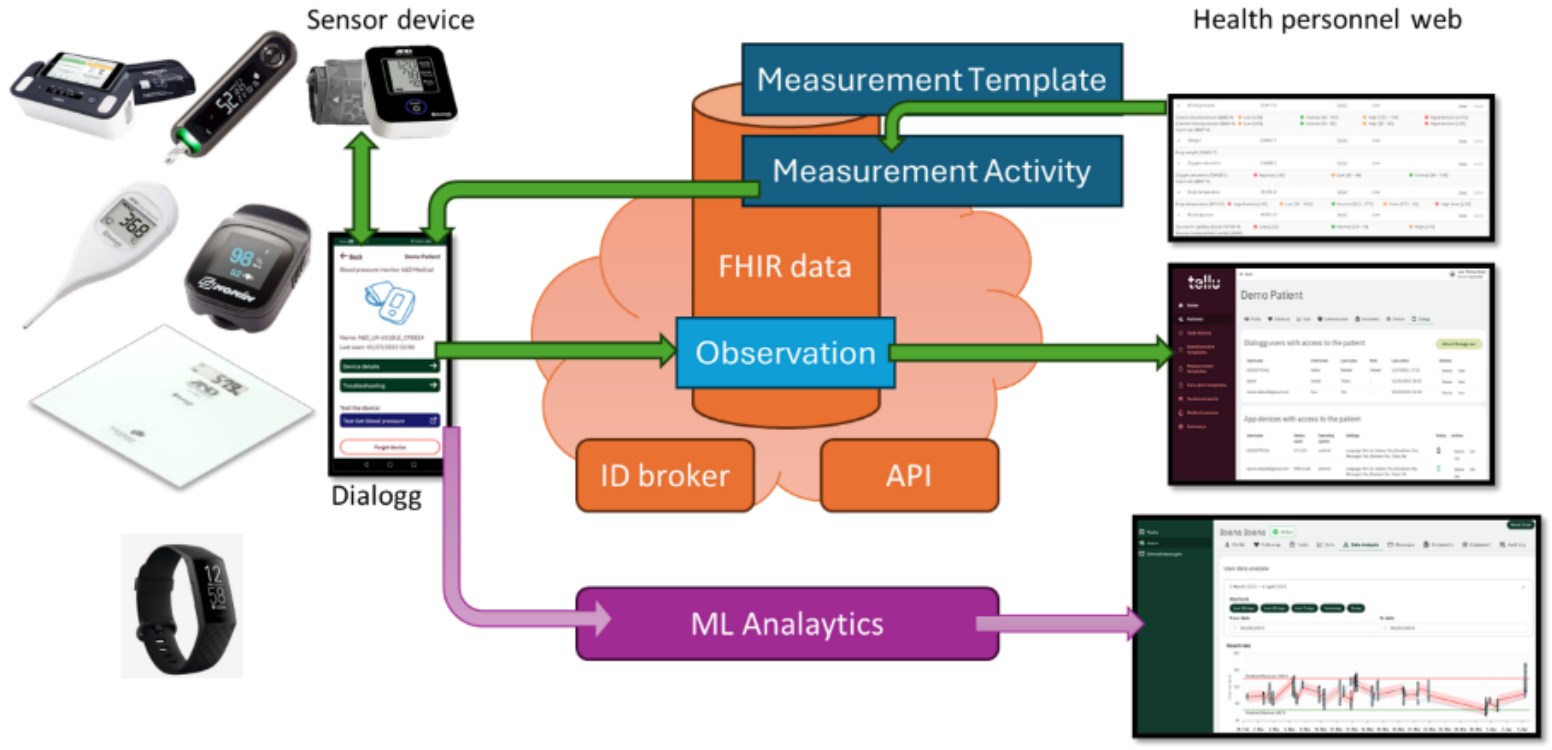

**Figure 5: Overview of the digital care management platform.**

The Digital Care Management Platform (**TelluCare**, see Figure 5) serves as the central hub for data aggregation, visualization, and communication. It supports HL7 FHIR standards for interoperability and includes secure APIs for integration with hospital systems and municipal healthcare services.

- **Remote Patient Monitoring (RPM):** Patients are equipped with CE-certified medical devices, including smart scales, blood pressure monitors, glucometers, pulse oximeters, and thermometers. These devices automatically transmit health data via Bluetooth to the platform.
- **Mobile Application (Dialogg):** Patients use the Dialogg application to report measurements, receive reminders, and access personalized health information. The application also facilitates asynchronous communication between patients and care teams.

- **Activity Tracking Devices:** A wearable smartwatch (Fitbit) continuously monitors physical activity, sleep patterns, heart rate, and other wellness indicators. These data provide longitudinal insights into patient health and behaviour over time.

Before discharge, patients are screened for eligibility, and those enrolled in the intervention are introduced to the TransCare technology ecosystem, including both medical and non-medical monitoring devices. Patients receive training in the use of the equipment and the Dialogg application, which serves as the primary interface for daily reporting and communication. Individual care plans are configured within the TelluCare platform, linking monitoring schedules with clinical oversight. Following discharge, patients begin using the devices in their home environment. Vital signs, activity levels, and other health indicators are collected and transmitted to the platform. Data captured through the Dialogg application include (i) intraday Fitbit time-series data, (ii) daily summary information, and (iii) data from other sensors and devices integrated into the RPM platform, such as blood pressure monitors and other connected medical devices. The platform visualizes health trends and generates alerts based on predefined thresholds. Healthcare professionals access dashboards to monitor patient status, review alerts, adjust care plans, and intervene when clinically appropriate.

The **Digital Assistant (Memas, see Figure 6)** provides patients with educational content and self-assessment tools designed to support health literacy, engagement in care, and self-management. Memas aims to promote safety, coping, independence, and overall quality of life. Within TransCare, the **Memas Rehabilitation** application supports individuals following a care plan after hospital discharge and throughout the rehabilitation process at home. Memas serves as both a communication aid and an information resource for patients, family members, and caregivers during the post-discharge period.

The platform includes two user roles with corresponding applications. **The Patient (end user)** is the primary user of the Memas application and receives support throughout the transitional care period following hospital discharge. The application provides:

- a calendar of appointments and activities;
- contact information for relevant healthcare providers and support persons;
- information related to the patient's diagnosis and care plan;
- personalized recommendations on diet and physical exercise;
- access to well-being and self-assessment questionnaires; and
- direct access to the Dialogg application for communication and monitoring.

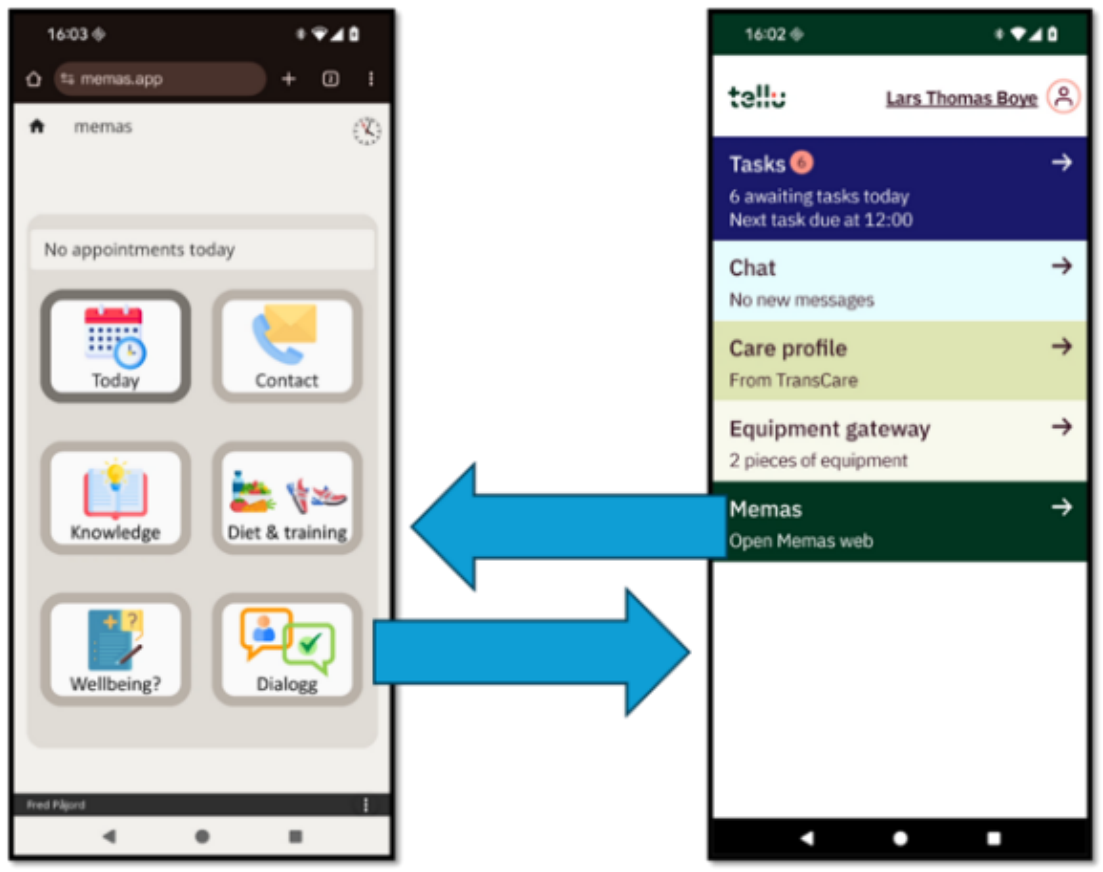

**Figure 6: Overview of the communication platform.**

**The Administrator** is a support person responsible for configuring, customizing, and maintaining the application for the patient. Administrators can create and manage multiple patient profiles, and more than one administrator can be assigned to the same patient when required.

**ML-Based Analytics (see Figure 7)** apply AI techniques to the analysis of daily activity and vital-sign data collected through the RPM platform. The aim is to generate clinically relevant insights that support proactive intervention and help healthcare professionals identify situations that may require further clinical assessment. In its current implementation, the component employs a personalized deep-learning model based on the Transformer architecture to predict heart rate during physical activities such as walking, running, swimming, and cycling [28]. By comparing predicted and observed heart rate values, the system can identify deviations that may indicate changes in a patient's condition and warrant clinical attention. A personalized modelling approach is used, allowing the model to adapt to each patient's physiological characteristics and patterns of activity. This recognizes the considerable variation in physiological responses between individuals and supports more accurate interpretation of patient data. The model operates on anonymized data and provides decision-support information to healthcare professionals through the TransCare platform. By analysing trends over time and identifying potential anomalies, the system contributes to more proactive monitoring and follow-up of patients during the transition from hospital to home.

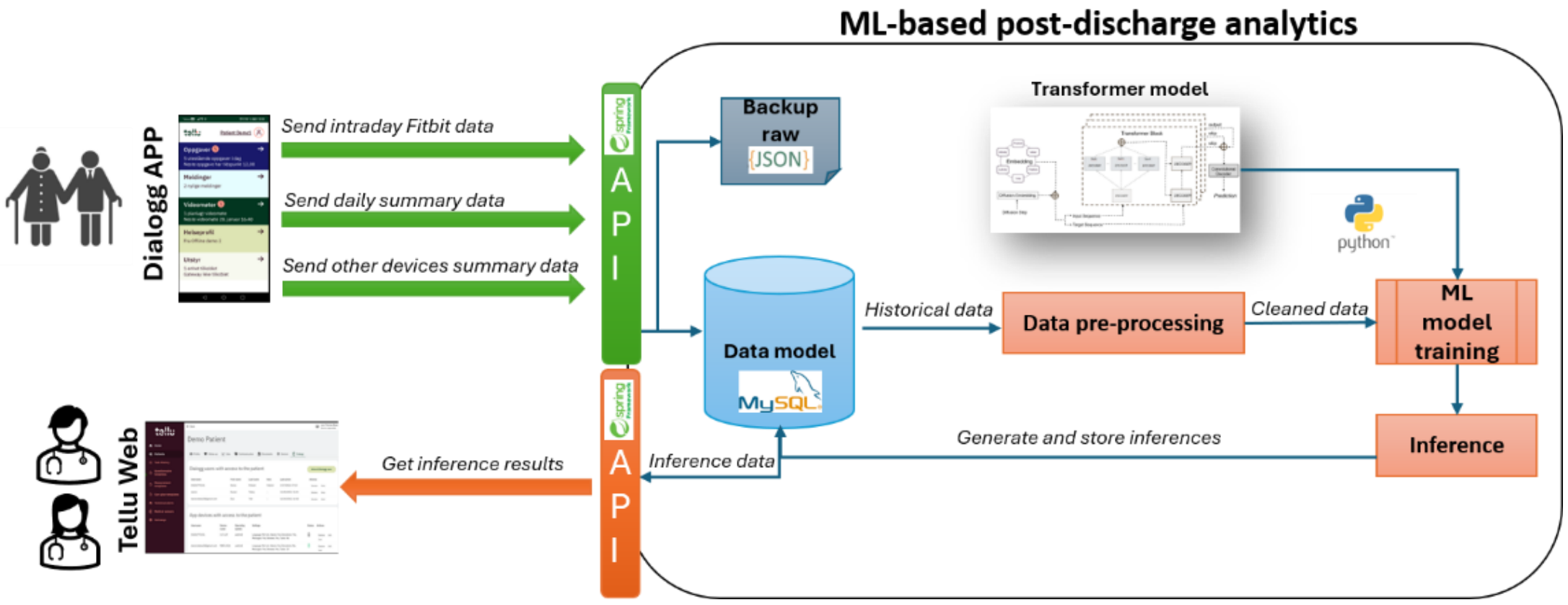

**Figure 7: Overview of the AI analytics integration**

## 3. Results

The redesigned care pathways in TransCare integrate digital technologies across multiple stages of the patient journey to support safer transitions from hospital to home, strengthen care coordination, and provide healthcare professionals with real-time clinical insights. Technology is a core component of the redesigned care model, supporting communication, monitoring, and decision-making throughout the care pathway. However, its implementation is carefully adapted to the specific needs, infrastructure, and digital maturity of each pilot site. In Italy, the focus is on older adults with frailty and multimorbidity. The platform is configured to support both patients and caregivers, with particular attention to accessibility for individuals with mild cognitive impairment through simplified user interfaces and localized educational content aligned with national healthcare standards. Continuous training is provided to promote effective and sustained use of the platform. In Romania, the emphasis is on cardiology patients at increased risk of readmission. The RPM system is integrated into existing hospital workflows, with cardiologists playing a central role in patient recruitment, clinical interpretation of monitoring data, and ongoing patient engagement. In Norway, the system is integrated within municipal healthcare services and is complemented by technologies such as camera supervision and personal alarm solutions to support patients receiving care at home, particularly in rural areas. Across all pilot sites, implementations are adapted to local organizational requirements and operational realities. The solutions comply with national regulations, language requirements, and data protection legislation, including the General Data Protection Regulation (GDPR), ensuring the secure and ethical management of patient data.

The redesigned matrix (Figure 8) in Italy aims to strengthen continuity of care and improve coordination between hospital services, patients, and home-based follow-up activities. The matrix retains the same macro-phases described in the original pathway while incorporating the interventions introduced through the TransCare project. The redesigned pathway clearly defines interactions between the key stakeholders involved in the care process, including patients, hospital professionals, and healthcare services responsible for post-discharge follow-up. Patients and caregivers receive training and support for home-monitoring activities during the first 90 days after discharge. This includes weekly telephone follow-up from TransCare project staff to support continued engagement and promote effective use of home-monitoring devices. Overall, the redesigned pathway provides a structured framework for improving transitional care at IRCCS-INRCA through enhanced communication, standardized assessments, patient empowerment, and the integration of home monitoring into the post-discharge follow-up process.

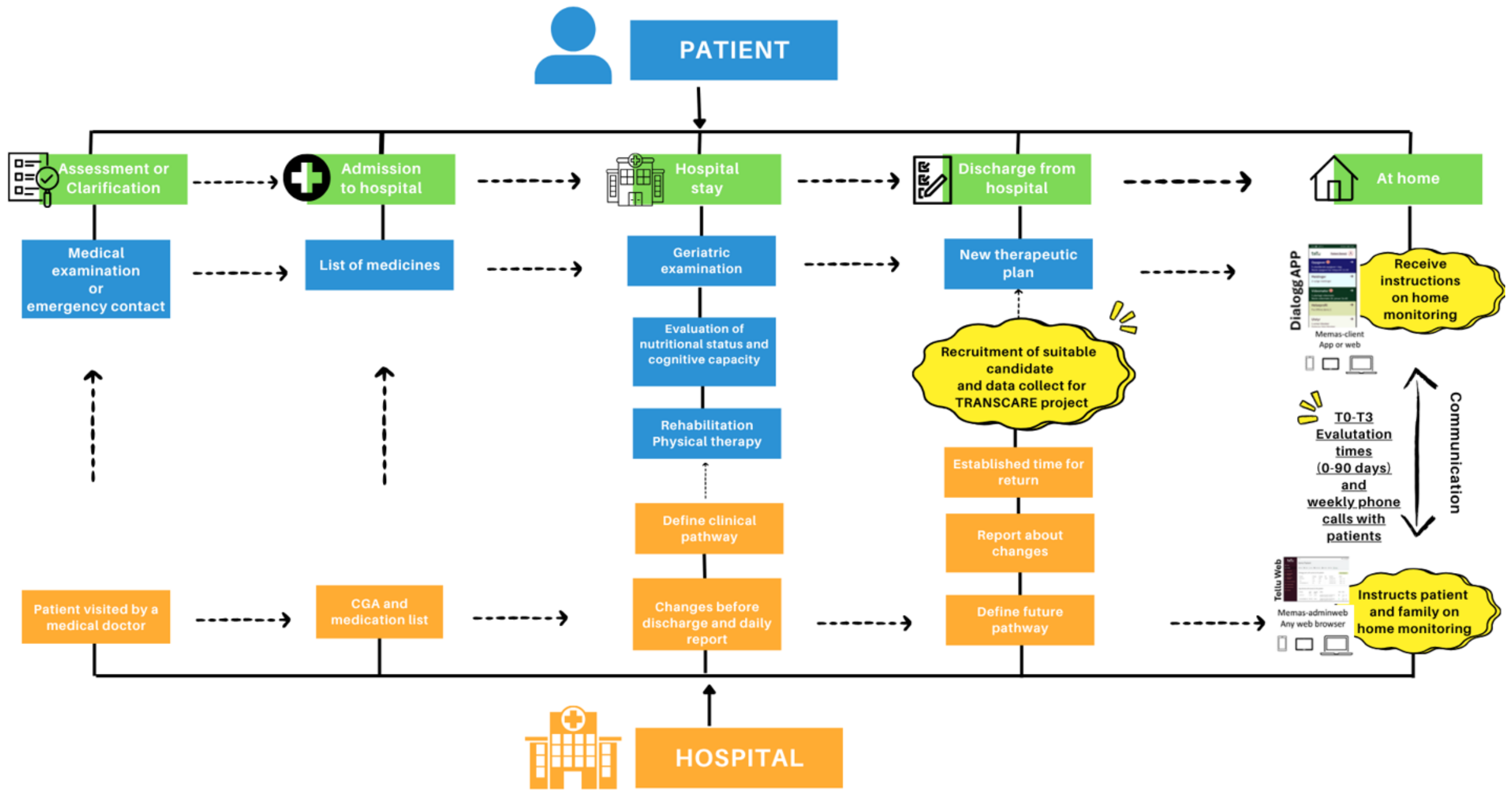

**Figure 8: Italy redesigned care pathway matrix**

The redesigned matrix (Figure 9) in Norway presents a more structured representation of the care pathway developed at FAR. The redesign aims to strengthen coordination between hospital and municipal healthcare services, clarify roles and responsibilities, and improve continuity of care throughout the transition from hospital to home. The revised pathway retains the same key phases: assessment and clarification, hospital admission, hospital stay, discharge and return home, and post-discharge follow-up. However, it introduces clearer points of interaction between stakeholders and more explicit coordination mechanisms across care settings. In particular, the role of the SPOC is further emphasized as a central coordinating function, facilitating timely information exchange and supporting seamless transitions between levels of care. A key enhancement in the redesigned pathway is the systematic integration of digital health and welfare technologies. Patients selected for the TransCare intervention are introduced to RPM solutions either before discharge or shortly after returning home and receive practical training in the use of monitoring devices and the Dialogg application. This helps ensure that both patients and caregivers are prepared to participate in health monitoring and self-management following discharge. During the post-discharge phase, the pathway incorporates structured follow-up through RPM, scheduled check-ins, and proactive responses to system-generated alerts. Municipal healthcare services play a central role in monitoring incoming data, responding to deviations from expected health parameters, and maintaining regular contact with patients. By enabling earlier identification of potential deterioration, the redesigned pathway supports timely intervention and may contribute to reducing the risk of hospital readmission.

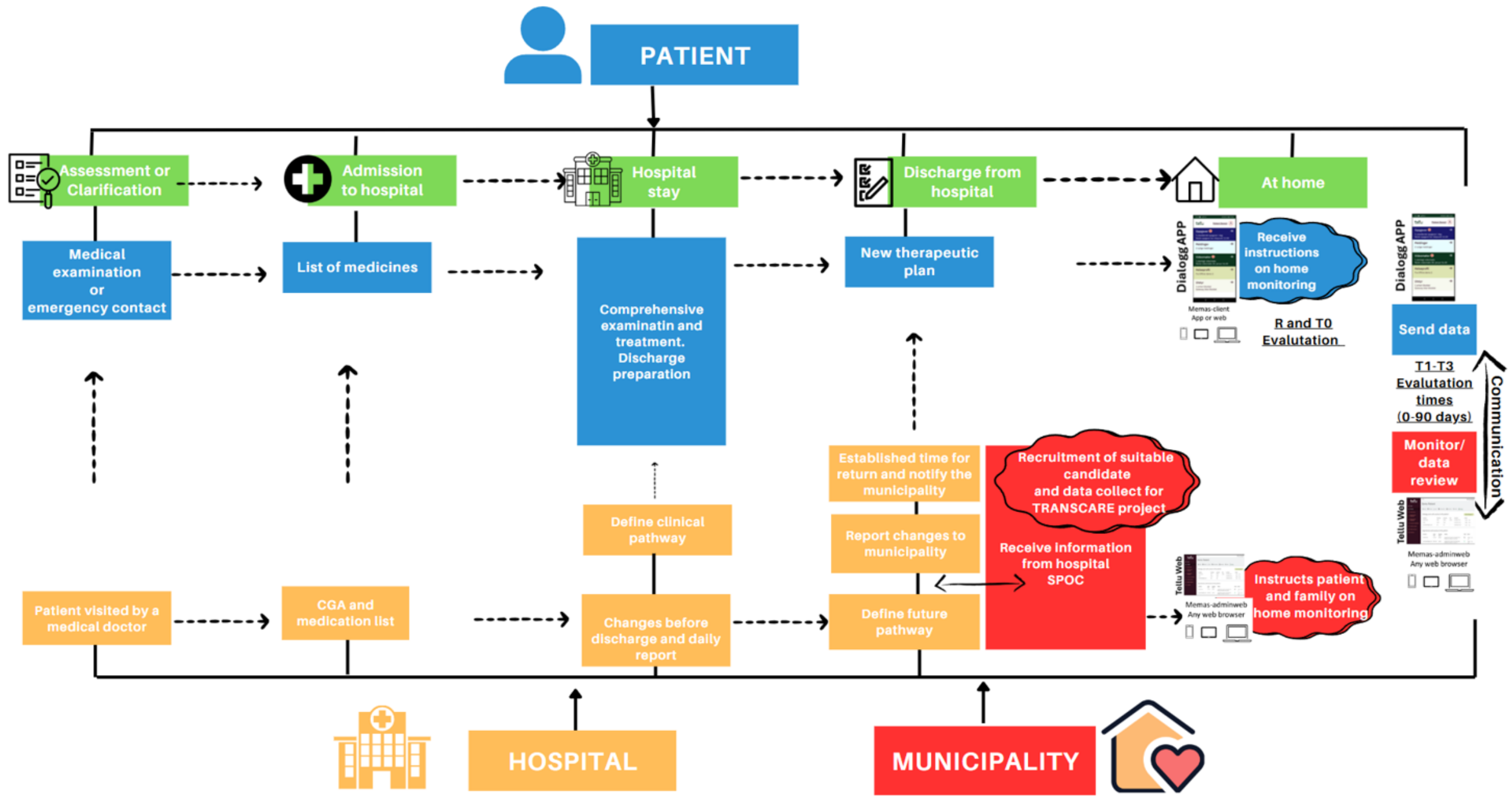

**Figure 9: Norway redesigned care pathway matrix**

The redesigned matrix for Romania (Figure 10) provides a structured overview of the patient pathway, from the onset of symptoms and initial assessment through hospital admission, discharge, and return home. The redesigned pathway highlights the integration of remote patient monitoring (RPM) into post-discharge care, with the aim of strengthening continuity of care and supporting coordination between patients and healthcare professionals following the transition from hospital to home. Patients are recruited during hospitalization once they are considered clinically ready for discharge. At this stage, patients and their families receive training in the use of home-monitoring devices and the transmission of clinical data, enabling them to actively participate in post-discharge monitoring and follow-up. A key feature of the redesigned pathway is the maintenance of continuous communication between patients and the clinical team after discharge. Patients transmit daily health data collected at home, which are reviewed by healthcare professionals. When clinically appropriate and feasible, treatment plans may be adjusted remotely based on the information received. This approach supports the early identification of changes in a patient's clinical status and promotes timely intervention when required. Overall, the redesigned pathway aims to reduce fragmentation of care by establishing a more continuous and coordinated follow-up process after discharge, thereby strengthening continuity of care and supporting safer transitions from hospital to home.

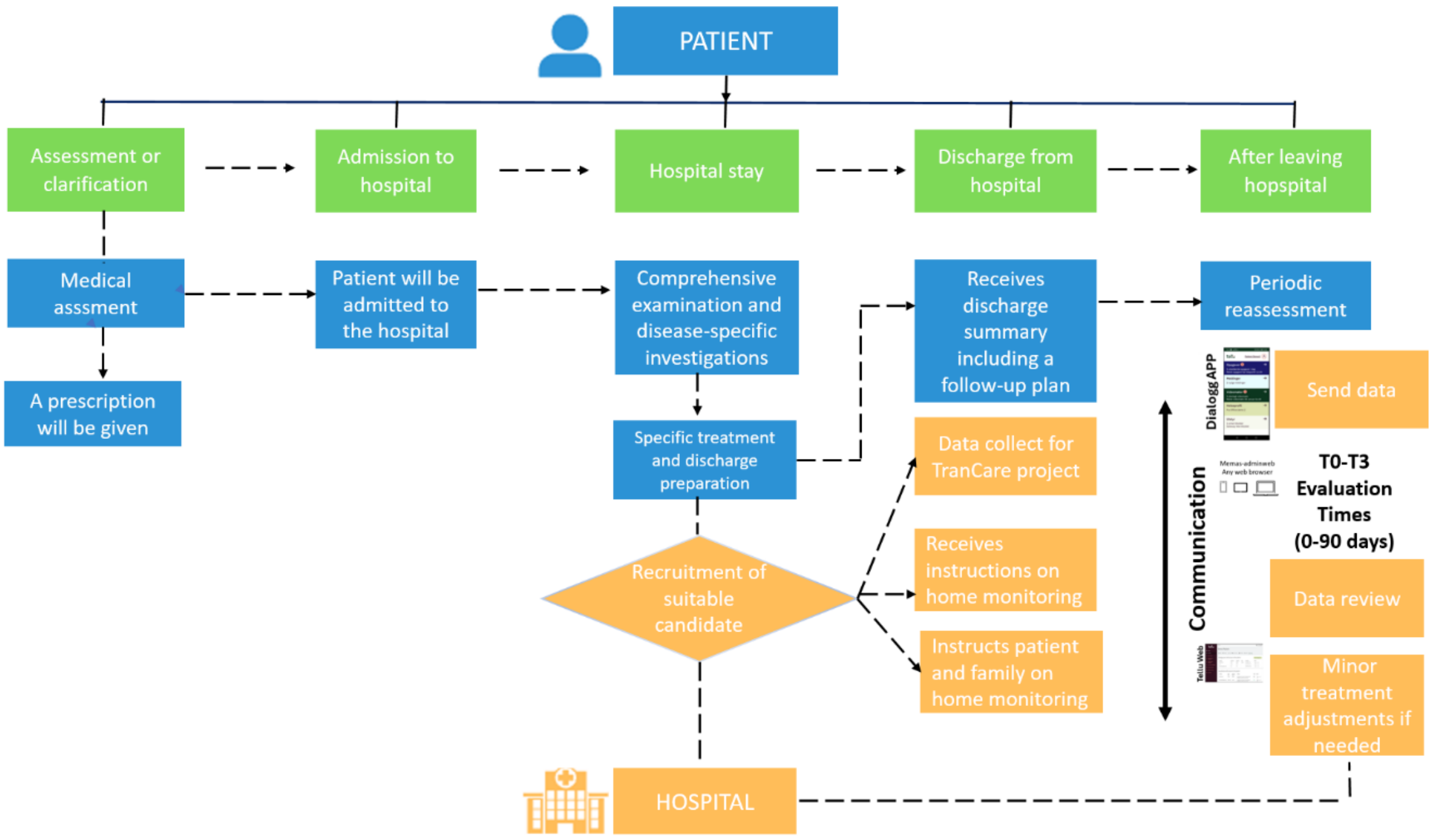

**Figure 10: Romania redesigned care pathway matrix**

## 4. Discussion

The development and early implementation of the redesigned transitional care pathways across the three pilot sites provide valuable insights into both shared challenges and context-specific differences. Based on project findings and iterative pathway refinements, the results suggest that improving transitional care requires a combination of organizational change, technology integration, and continuous stakeholder engagement.

A key finding across all sites is the presence of fragmentation in communication and coordination between levels of care, particularly during discharge and post-discharge follow-up. While the underlying causes differ, ranging from limited integration between hospital and community services (INRCA), to resource constraints and high patient volumes (HINS), to coordination and interoperability challenges across municipal services (FAR), the consequences are similar: reduced continuity of care and an increased risk of adverse outcomes, including hospital readmission. These shared challenges highlight the importance of designing pathways that not only define clinical activities but also explicitly structure information exchange, responsibilities, and follow-up processes.

At the same time, important differences between the pilot sites influenced how the redesigned pathways were adapted and implemented. INRCA operates within a geriatric context characterized by frailty, multimorbidity, and limited integration with community services, emphasizing the need for strengthened post-discharge coordination and caregiver involvement. HINS, as a high-capacity cardiology center, faces challenges related to high patient volumes and limited resources available for follow-up, making efficiency, prioritization, and RPM particularly relevant. In contrast, FAR represents a decentralized, municipality-based healthcare system where coordination across services and geographical distances are central challenges, and where the role of the SPOC is particularly important. These differences demonstrate that transitional care models must remain flexible and adaptable rather than standardized if they are to remain relevant and feasible across diverse healthcare systems.

A major lesson learned is that the integration of digital technologies, including RPM, digital communication platforms, and AI-supported analytics, can strengthen continuity of care when accompanied by clearly defined workflows, role clarification, and adequate training. Technology alone does not resolve coordination challenges; rather, it acts as an enabler within a well-structured care pathway. Early implementation experiences underline the importance of preparing patients and caregivers before discharge, supporting healthcare professionals in the adoption of new tools, and aligning digital solutions with existing organizational practices and capacities.

Another important insight is the value of the co-design approach adopted throughout the project. The active involvement of patients, caregivers, clinicians, and technology partners contributed to pathways that are more realistic, acceptable, and aligned with stakeholder needs.

This approach facilitated the identification of practical barriers, increased stakeholder ownership, and supported the iterative refinement of both care pathways and technologies. The findings also highlight that successful implementation depends not only on technical and clinical validity, but also on user engagement, organizational readiness, and sustained stakeholder commitment.

Positive outcomes observed during the redesign process include greater clarity of roles and responsibilities, more structured discharge planning, and enhanced follow-up supported by RPM and proactive communication. The introduction of continuous monitoring and data-driven insights creates opportunities for earlier identification of deterioration, more timely intervention, and a more personalized approach to care. Furthermore, increased involvement of patients and family members may contribute to empowerment and improved adherence to care plans.

However, the study also confirms that transitional care redesign is a complex and iterative process. Given the early stage of implementation, many findings remain preliminary, and additional evaluation is required to assess effectiveness, long-term impact, scalability, and sustainability. Organizational constraints, resource limitations, and varying levels of digital maturity continue to influence implementation and require ongoing adaptation. At the Italian pilot site, the vulnerability and frailty of the target population also present practical challenges in identifying patients who both meet the inclusion criteria and possess sufficient independence to use digital monitoring technologies effectively.

Several limitations should be considered when interpreting these findings. First, the redesigned pathways are currently in the early stages of implementation, and comprehensive outcome data are not yet available. Second, the pilot sites operate within substantially different healthcare systems, organizational structures, and patient populations, which may limit the generalizability of the findings. Third, implementation experiences have been influenced by local resource availability, workforce capacity, and digital maturity. Finally, the study primarily reports pathway redesign and early implementation experiences rather than clinical effectiveness outcomes.

In summary, the TransCare experience demonstrates that improving transitional care requires a balanced integration of process redesign, technology, and stakeholder collaboration. The iterative and context-sensitive approach adopted in this project provides a valuable foundation for continued refinement and offers transferable lessons for healthcare systems seeking to strengthen transitions from hospital to home.

## 5. Conclusion

The TransCare experience demonstrates that effective hospital-to-home transitions require the integration of care pathway redesign, digital technologies, and stakeholder collaboration. Across the three pilot sites, digitally supported pathways addressed fragmentation in communication, coordination, discharge planning, and post-discharge follow-up while remaining adaptable to different healthcare contexts. RPM, digital communication tools, and AI-supported analytics have the potential to strengthen continuity of care and support earlier intervention when embedded within clearly defined workflows and supported by appropriate training and organizational readiness. The co-design and iterative approach facilitated the adaptation of both care processes and technologies to the needs of patients, caregivers, and healthcare professionals. Although further evaluation is required to assess long-term outcomes and scalability, the findings provide transferable lessons for the development of context-sensitive, technology-enabled transitional care pathways across European healthcare systems. Future work will focus on pilot testing the redesigned pathways in real-world settings and evaluating their feasibility, usability, and impact on transitional care outcomes and patient experience.

**Author Contributions:** Conceptualization: CG, OA, RB, IA, TC; Investigation: LR, EB, MO, AN, RH, LTB; Methodology: IA, TC, CG, OA, RB; Visualization: EB, MO, AS, TG; Writing-original draft: IA, TC, LR, EB, MO, CG, TG, LTB, OA, AN; Writing-review & editing: IA, RH, AS, CG, OA, RB.

## Statements and Declarations

**Guarantor:** CG.

**Ethical considerations**: In Italy the study was approved by the Marche Territorial Ethics Committee (CET MARCHE) (Study Ref. n. 102/2025 approved 19/06/25). In Romania, it was approved by the hospital's Ethics Committee under approval no. 4678, dated 11 April 2024. In Norway, ethical approval was not required because the project was considered outside the scope of health research requiring review by the Medical Research Ethics Committee. The study protocol was recorded in ClinicalTrials.gov on the number NCT07731126.

**Declaration of conflicting interests:** The authors declared no potential conflicts of interest with respect to the research, authorship, and/or publication of this article.

**Funding:** This work was supported by a grant of the Ministry of Research, Innovation and Digitization, CNCS/CCCDI - UEFISCDI, project number ERANET-PARTENERIAT-THCS-TransCare-1, within PNCDI IV. It was also supported by the Research Council of Norway (NO), and Italian Ministry of Health (IT) under the Transforming Health & Care Systems (THCS) Programme. Views and opinions expressed are however those of the author(s) only and do not necessarily reflect those of the European Union, the National Agencies or THCS.

**Data availability:** The data analyzed during the current study is not publicly available. Access is restricted due to the sensitive nature of the data collected from participants, the terms of the informed consent process, and confidentiality obligations under the project consortium agreement and applicable data protection legislation.